\documentstyle[prd,aps,floats,twocolumn]{revtex}
\preprint{DAMTP-1999-182}
\date{Submitted to Phys. Rev. D, 22/12/1999; revised 28/4/2000, 8/8/2000}
\tighten
\begin{document}
\draft
\def\sqr#1#2{{\vcenter{\hrule height.3pt
      \hbox{\vrule width.3pt height#2pt  \kern#1pt
         \vrule width.3pt}  \hrule height.3pt}}}
\def\square{\mathchoice{\sqr67\,}{\sqr67\,}\sqr{3}{3.5}\sqr{3}{3.5}}
\def\today{\ifcase\month\or
  January\or February\or March\or April\or May\or June\or
  July\or August\or September\or October\or November\or December\fi
  \space\number\day, \number\year}
\input epsf
%%%%%%%%%%%%%%%%%%%%%%%%%%%%%%%%%%%%%%%%%%%%%%%%%%%%%%%%%%%%%%%%%%%%%%%%%%

\title{Topological defects: fossils of an anisotropic era?}

\author{P. P. Avelino${}^{1,2}$\thanks{
Electronic address: pedro\,@\,astro.up.pt} 
and C. J. A. P. Martins${}^{3}$\thanks{Also at C.A.U.P.,
Rua das Estrelas s/n, 4150 Porto, Portugal.
Electronic address: C.J.A.P.Martins\,@\,damtp.cam.ac.uk}}

\address{${}^1$ Centro de Astrof\'{\i}sica, Universidade do Porto\\
Rua das Estrelas s/n, 4150 Porto, Portugal}

\address{${}^2$ Dep. de F{\' \i}sica da Faculdade de Ci\^encias da Univ. do Porto\\
Rua do Campo Alegre 687, 4169-007 Porto, Portugal}

\address{${}^3$ Department of Applied Mathematics and Theoretical Physics\\
Centre for Mathematical Sciences, University of Cambridge\\
Wilberforce Road, Cambridge CB3 0WA, U.K.}

\maketitle
\begin{abstract}

{We consider the evolution of domain walls produced during an 
anisotropic phase in the very early universe, showing that the resulting
network can be very anisotropic. If the domain walls are produced during an 
inflationary era, the network will soon freeze out in comoving 
coordinates retaining the imprints of the anisotropic regime, even though
inflation makes the universe isotropic. Only at late times, when the typical
size of the major axis of the domain walls becomes smaller
than the Hubble radius, does the network evolve rapidly towards isotropy.

Hence, we may 
hope to see imprints of the anisotropic era if by today the typical 
size of the major axis of the domain walls is of the order of the 
Hubble radius, or if the walls re-entered it only very recently. 
Depending on the mass scale of the domain walls,
there is also the possibility that they re-enter at earlier times,
but their evolution remains 
friction-dominated until recently, in which case the signatures of the
anisotropic era will be much better preserved.
These effects are expected to occur in generic topological defect models.}

\end{abstract} 
\newpage

%%%%%%%%%%%%%%%%%%%%%%%%%%%%%%%%%%%%%%%%%%%%%%%%%%%%%%%%%%%%%%%%%%%%%%%%%%
\section{Introduction}
\label{intro} 

It is well know that the `Hot Big Bang' model \cite{kolb},
despite its numerous successes,
is plagued by a number of `initial conditions' problems, of which the
horizon, flatness and unwanted relic ones are the best known. The standard
way to solve them is to invoke an epoch of cosmological
inflation \cite{guth,linde,lyth}, a relatively brief period of exponential
(or quasi-exponential) cosmological expansion.
The way inflation solves these problems is, loosely speaking, by erasing
all traces of earlier epochs and re-setting the universe to a rather simple
state. Indeed, inflation is so efficient in this task that a number of
people have wondered if one can ever hope to probe the physics of a
pre-inflationary epoch.

There are, however, a small number of possible pre-inflationary relics.
For example, the recent work by Turok and collaborators \cite{curv}
shows that curvature
can, in some sense, survive inflation. Another class of inflationary
survivors are topological defects \cite{vsh}, formed at phase transitions
either before or during inflation \cite{shafi,yoko,hodges}).
It is known (see, e.g. \cite{kolb,linde,acm2}) that one needs
about 20 e-foldings of inflation\footnote{The exact number is of course
model-dependent.} to solve the monopole problem. One can reverse the argument
and say that monopoles can survive about 20 e-foldings of inflation. The
inflationary epoch itself will obviously push the monopoles outside the horizon,
but the subsequent evolution of the universe tends to make them come back
inside, so if the inflationary epoch is not too long they can still have
important cosmological consequences.

Cosmic strings are even more successful, being able to survive about 50
e-foldings. The reason for this difference is that their non-trivial
dynamics \cite{ms1,ms2,ms3,acm2} makes them come back inside the
horizon faster than one might
naively have expected. The above two numbers are typical, but there are
specific models where defects can survive even longer. One example is that
of open inflation scenarios \cite{bgt}. In this case the universe undergoes two
different inflationary epochs---roughly speaking, a period of `old inflation'
followed by one of `new inflation'. As pointed out by Vilenkin \cite{vil},
one can
expect that defects will form between the two inflationary epochs. In this
case, a collaboration \cite{acm2} including the present authors
has recently shown that
not only will cosmic strings survive the entire second inflationary epoch,
regardless of how long it lasts\footnote{Note that in these models the duration
of the second inflationary epoch is fixed by the present value of the density
of the universe \cite{bgt}.}, but they will in fact be
back inside the horizon by the
time of equal matter and radiation densities. In such models, monopoles can
survive up to about 30 e-foldings.

Now, given that defects seem to be so successful surviving inflation, and that
one expects them to be frozen out while they are outside the horizon, one
can think of a further interesting possibility. For the best-studied case of
cosmic strings, it is well known that the scaling properties of the network
depend on the background cosmology \cite{ms1,ms2,ms3}. Moreover, in some
cases (typically when
their evolution is friction-dominated) they can retain a `memory' of the
initial conditions, or the general properties of the cosmology in which
they find themselves at early times, for quite a large number of orders of
magnitude in time \cite{ms1,ms3}. It is therefore conceivable
that if such an imprint of
an early cosmological epoch is retained by a defect network which manages to
survive inflation, we might still be able to observe it today.

We believe that this is a general feature of defect models, and a number
of non-trivial pieces of information about the very early universe can
probably be preserved in this way. In the present
paper we will restrict ourselves to a simple example. We will discuss
the possibility of a domain wall network retaining information about an
early anisotropic phase of the universe.
There are very strong constraints \cite{zeld} on the mass of domain walls
formed after inflation, due to the fact that their density decays more
slowly than the radiation and matter densities. However, these can be evaded
by walls forming before or during inflation.
In a subsequent paper, we will discuss the more interesting, but also more
complicated, case of cosmic strings.

The plan of the paper is as follows. In section \ref{evequations} we briefly
describe our background (Bianchi I) cosmology and the basic evolutionary
properties of the
domain walls. In particular, we focus on the approach to isotropy during
inflation, which is discussed through both analytic arguments and numerical
simulations. We emphasise that these simulations {\it do not} include
the defects. However, they serve an important purpose, as they are
used in the subsequent discussion to show that the timescale needed for
isotropization is compatible with the `survival' on anisotropic defect networks.

We provide a description of our numerical simulations of
domain wall evolution in section \ref{nums}. These are analogous to those
of Press, Ryden \& Spergel \cite{PRS}, and the interested reader is referred
to this paper for a more detailed discussion of some relevant numerical
issues. Here defect networks are evolved in an {\it isotropic},
matter-dominated
(ie, post-inflationary universe), and their main purpose is to show that
isotropic and anisotropic networks will evolve in different ways, so
two such networks can in principle be observationally distinguished as
they re-enter the horizon. Our main results are presented and discussed
in section \ref{redi},
and finally we present our conclusions and discuss future work
in section \ref{conc}.

Throughout this paper we will use fundamental units in which $c=1$.

%%%%%%%%%%%%%%%%%%%%%%%%%%%%%%%%%%%%%%%%%%%%%%%%%%%%%%%%%%%%%%%%%%%%%%%%%%
\section{Evolution equations for domain walls}
\label{evequations}

We consider the evolution of a network of domain walls in a k=0 anisotropic 
universe of Bianchi type I with line element \cite{anis}:
\begin{equation}
ds^2=dt^2-X^2(t)dx^2-Y^2(t)dy^2-Z^2(t)dz^2
\label{metric}
\end{equation}
where $X(t)$,$Y(t)$ and $Z(t)$ are the cosmological expansion factors 
in the $x$, $y$ and $z$ directions respectively, and $t$ is the physical 
time. The dynamics of a scalar filed $\phi$ is 
determined by the Lagrangian density,
\begin{equation}
{\cal L}=-{1 \over {4 \pi}}\left({1 \over 2} \phi_{,\alpha} \phi^{, \alpha} + 
V(\phi)\right),
\label{action1}
\end{equation}
where we will take $V(\phi)$ to be the generic $\phi^4$ potential with two 
degenerate minima given by
\begin{equation}
V(\phi)=V_0\left({\phi^2 \over \phi_0^2}-1\right)^2.
\label{potential}
\end{equation}
This obviously admits domain wall solutions \cite{vsh}.
By varying the action 
\begin{equation}
S=\int dt \int d^3x {\sqrt {-g}} {\cal L},
\label{action2}
\end{equation}
with respect to $\phi$ we obtain the field equation of 
motion:
\begin{equation}
{{\partial^2 \phi} \over {\partial t^2}} + \theta
{{\partial \phi} \over {\partial t}}
 - {\nabla'}^2 \phi=
-{{\partial V} \over {\partial \phi}}.
\label{dynamics}
\end{equation}
where 
\begin{equation}
{\nabla'}^2={1 \over X^2}{{\partial^2} \over {\partial x^2}}
+{1 \over Y^2}{{\partial^2} \over {\partial y^2}}
+{1 \over Z^2}{{\partial^2} \over {\partial z^2}},
\label{laplacian}
\end{equation}
with $\theta(t)={{\dot W} / W}$ and $W(t)=XYZ$. The dynamics of 
the universe is described by the Einstein field equations. Here we 
shall seek perfect fluid solutions. The time component of the 
Einstein equation then becomes
\begin{equation}
{\dot \theta} + A^2 + B^2 + C^2 = - {1 \over 2} k (\rho +3p),
\label{einstein0}
\end{equation}
while the spatial components give
\begin{equation}
{\dot A} + \theta A = {\dot B} + \theta B = {\dot C} + \theta C = 
{1 \over 2} k (\rho -p),
\label{einsteini}
\end{equation}
with $A={\dot X}/X$, $B={\dot Y} / Y$ and $C={\dot Z}/Z$, $\theta=A+B+C$ and 
$k=8 \pi G/c^2$ and $i=1,2,3$. It is straightforward to combine 
equations (\ref{einstein0},\ref{einsteini}) to obtain:
\begin{equation}
AB+BC+CA=k \rho.
\label{einconst}
\end{equation}
In the following discussion we will 
make the simplification that $X(t)=Z(t)$ (and therefore $A=C$) 
and consider the dynamics of the 
universe during an inflationary phase with $\rho=-p= const$. 
In this 
case $H^2 \equiv k \rho/3 = const$ and the Einstein field 
equations (\ref{einstein0},\ref{einsteini},\ref{einconst}) imply:
\begin{equation}
{\dot A} +{3 \over 2}(A^2 - H^2)=0,
\label{dynA}
\end{equation}
while $B$ can be found from the suggestive relation
\begin{equation}
\frac{B}{A} = \frac{1}{2}\left(\frac{3H^2}{A^2}-1\right).
\label{dynB}
\end{equation}
Equation (\ref{dynA}) has two solutions, depending on the initial conditions.
If $A_i<H$, then $A$ is the smaller of the two dimensions and the shape of
spatial hyper-surfaces is similar to that of a rugby ball. Then the solution
is
\begin{equation}
\frac{A}{H}=\tanh\left[\frac{3}{2}H(t-t_i)+\tanh^{-1}\left(\frac{A_i}{H}\right)\right],
\label{soldynAsmall}
\end{equation}
with $A_i=A(t_i)$.
On the other hand, if $A_i>H$, then $A$ is the larger of the dimensions and
the shape of spatial hyper-surfaces is similar to that of a pumpkin. Then the
solution is
\begin{equation}
\frac{A}{H}=\coth\left[\frac{3}{2}H(t-t_i)+\coth^{-1}\left(\frac{A_i}{H}\right)\right]\, .
\label{soldynAlarge}
\end{equation}
Note that in both cases the ratio $A/H$ tends to unity exponentially fast, and
hence the same happens with the ratio $B/A$. In other words, inflation tends to
make the universe more isotropic, as expected. An easy way to see
this is to consider the
ratio of the two different dimensions, $D=B/A$, and to study its evolution
equation. One easily finds
\begin{equation}
{\dot D}=\sqrt{6}H\left(D+\frac{1}{2}\right)^{1/2}(1-D),
\label{evratio}
\end{equation}
which has an obvious attractor at $D=1$.  

Note that even though we have so far assumed (for simplicity)
that $p=-\rho$, the same analysis can be carried out 
for an inflating universe with $p=(\gamma-1) \rho$ with $\gamma \neq 0$ by 
numerically solving the conservation equation
\begin{equation}
{\dot \rho} + \theta (\rho+p)=0,
\label{soldynA1}
\end{equation}
together with equations (\ref{einsteini}) and (\ref{einconst}). Indeed, the
more general case will be relevant for what follows.

\begin{figure*}[t]
\begin{center}
\leavevmode \epsfbox{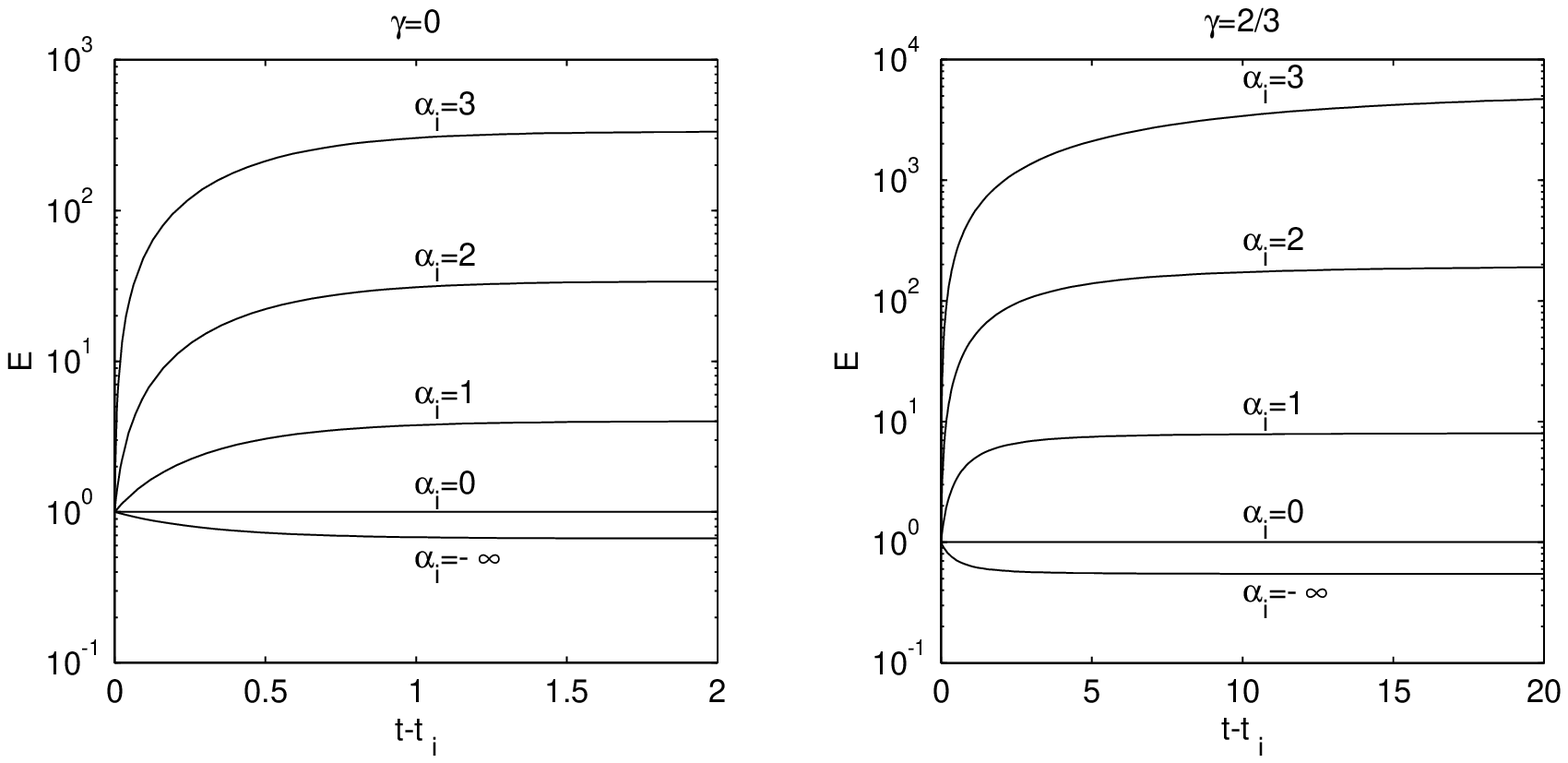}\\
\end{center}
\vskip-0.5in
\caption[]{Evolution of the asymmetry parameter $E=Y(t)/X(t)$ (according 
to eqns. (\ref{einstein0}) and 
(\ref{einsteini})) for 
several values of $\alpha_i =  \log_{10} (D_i)$ 
assuming $\gamma=0$ (left panel)
and $\gamma=2/3$ (right panel) respectively. Note that time is
given in units of $H^{-1}(t_i)$.}
\vskip0.5in
\label{figure1}
\end{figure*}

In figure \ref{figure1} we plot the evolution of the asymmetry 
parameter $E=Y(t)/X(t)$, according to eqns. (\ref{einstein0}) and 
(\ref{einsteini}), for 
several values of $\alpha_i =  \log_{10} (D_i)$ assuming $\gamma=0$ and 
$\gamma=2/3$ (note that $\gamma = 2/3$ is the maximum value of $\gamma$ which 
violates the strong energy condition). Note that we do not include the
defect network in the simulation. (We assume that the network at the 
initial time $t_i$ is statistically isotropic.) We take 
$X(t_i)=Y(t_i)$. We can see that depending on the initial degree of 
anisotropy, specified by $\alpha_i$, the value of $E$ can grow to be very 
large, especially if $\alpha_i$ is large. Moreover, although for 
$\gamma=0$, the 
value of $E$ becomes approximately constant in one Hubble time that 
does not happen so rapidly for inflating universes with larger $\gamma$. 
This removes the necessity of producing the domain walls right at the onset 
of the inflationary era.

What about the evolution of the domain walls? Based on rather general grounds,
we expect it to have a number of similarities with the much better studied
case of cosmic strings \cite{vsh}. In particular, one can define a
`characteristic length-scale', which we shall denote by $L$, that can be
roughly interpreted as a typical curvature radius or a correlation length of
the wall network. It is also a length-scale that measures the total energy of
the domain wall network per unit volume, since we can define
\begin{equation}
\rho\equiv\frac{\sigma}{L}\, ,
\label{defl}
\end{equation}
where $\sigma$ is the domain wall energy per unit area. Note that in a
more rigorous treatment that allowed for the expected build-up of
small-scale `wiggles' on the walls (in analogy with what happens for the
case of cosmic strings \cite{ms3}) each of these three length scales would be
different. However, for our present purposes it is adequate to suppose
that they are all similar.

Then we can expect to find two different evolution regimes. While the network
is non-relativistic, we expect it to be conformally stretched by the
cosmological expansion, and hence
\begin{equation}
L\propto a\, , \qquad \rho_w\propto a^{-1}\, .
\label{stretch}
\end{equation}
In this case there is essentially no dynamics.
An extreme example of this regime happens during inflation
We can see from eqn. 
(\ref{dynamics}) that due to the very rapid 
expansion which occurs in the inflationary regime the time derivatives 
of the field $\phi$ rapidly approach zero so that the network of domain 
walls will simply be frozen in comoving coordinates.

On the other hand, once the network becomes relativistic, one expects it
to evolve in a linear scaling regime where
\begin{equation}
L\propto t\, , \qquad \rho_w\propto t^{-1}\, .
\label{scaling}
\end{equation}
This is the case of `maximal' dynamics, in the sense that the network is
evolving (in particular, losing energy by wall collisions and re-connections)
as fast as allowed by causality.
We note that previous work of Press, Ryder and Spergel \cite{PRS}
suggests that there may be logarithmic corrections to this linear regime. 

%%%%%%%%%%%%%%%%%%%%%%%%%%%%%%%%%%%%%%%%%%%%%%%%%%%%%%%%%%%%%%%%%%%%%%%%%%
\section{Numerical simulations}
\label{nums}

At late times (after the inflationary epoch) the universe is homogeneous 
and isotropic with $A=B=C$ with the average dynamics of the universe being 
specified by the evolution of the scale-factor $a(t)$.
We now consider the evolution of isotropic and anisotropic defect networks
in this background. In particular, we are interested in determining
how the networks evolve as they re-enter the horizon, since if one
finds differences in the dynamics of the two cases then this should
translate into observational tests that will allow us to discriminate between
then and hence probe pre-inflationary physics.
 
It is useful for numerical purposes to 
re-write equation (\ref{dynamics}) as a function of the conformal time $\eta$ 
defined by $d \eta = dt/a$. In this case equation (\ref{dynamics}) becomes
\begin{equation}
{{\partial^2 \phi} \over {\partial \eta^2}} + 2 \frac{\dot a}{a}
{{\partial \phi} \over {\partial \eta}}
 - {\nabla}^2 \phi=
-a^{2}{{\partial V} \over {\partial \phi}}.
\label{dynamics1}
\end{equation}
with
\begin{equation}
{\nabla}^2={{\partial^2} \over {\partial x^2}}
+{{\partial^2} \over {\partial y^2}}
+{{\partial^2} \over {\partial z^2}}.
\label{laplacian1}
\end{equation}
When making numerical simulations of the evolution of domain wall networks
(or indeed other defects) it is
also often convenient to modify the equation of motion for the scalar field 
$\phi$ in such a way that the comoving thickness of the walls is fixed in 
comoving coordinates. This is known as the PRS algorithm \cite{PRS}, and it is
generally believed not to significantly affect the large-scale dynamics of 
domain walls.

We note, however, that recent high-resolution
simulations \cite{eps} have revealed that the accuracy of this algorithm is
not as good as has been claimed. This effect is expected to increase with
increasing dynamic range.
In particular, the PRS algorithm artificially
prevents the build-up of small-scale features on the domain walls (or, for that
matter, any other defect). This turns out to be crucial for a quantitatively
accurate description of their evolution, and hence for a reliable analysis
of their observational consequences. For our purposes in the present work,
however, the PRS algorithm is enough as an approximation to the true wall
dynamics. In a subsequent, more detailed publication we shall compare results
obtained using this algorithm with those from the true wall dynamics.

Having clarified this point, we will modify the evolution equation for the
scalar field  $\phi$ in the isotropic phase according to the PRS
prescription:
\begin{equation}
{{\partial^2 \phi} \over {\partial \eta^2}} + \beta_1 \frac{\dot a}{a}
{{\partial \phi} \over {\partial \eta}}
 - {\nabla}^2 \phi=
-a^{\beta_2}{{\partial V} \over {\partial \phi}}.
\label{dynamics2}
\end{equation}
where $\beta_1$ and $\beta_2$ are constants. 
We choose $\beta_2=0$ in order 
for the walls to have constant comoving thickness and $\beta_1=3$ by requiring 
that the momentum conservation law for how a wall slows down in an expanding 
universe is maintained \cite{PRS}.

We perform two-dimensional simulations of domain wall evolution for which 
$\partial^2 \phi / \partial z^2 =0$. 
These have the advantage of allowing 
a larger dynamic range and better resolution than tree-dimensional simulations.

We solve equation (\ref{dynamics2}) numerically assuming a matter-dominated 
Einstein-de-Sitter cosmology with $a \propto \eta^2$. We used a standard 
difference scheme second-order accurate in space and time 
and periodic boundary conditions (see \cite{PRS}
for a more detailed description of the algorithm and other
related numerical issues).

The initial properties of the network of domain walls depend strongly 
on the details of the phase transition which originated them. It is conceivable 
that the initial network is already formed asymmetric with the walls 
being elongated along preferred directions. However, this is beyond the
scope of the present paper. For our present purposes, we can
ignore this possibility and assume that the initial domain 
wall network is statistically isotropic. This assumption will not modify 
the conclusions of the paper---if anything, any {\it ab initio} anisotropies
would only enhance the effects we are describing.

Hence, we assume the initial 
value of $\phi$ to be a random variable between $-\phi_0$ and $\phi_0$ and the 
initial value of ${\dot \phi}$ to be equal to zero everywhere. We normalise 
the numerical simulations so that $\phi_0=1$. We set the conformal time 
at the start of the simulation and the comoving spacing between 
the mesh points to be respectively $\eta_i=1$ and $\Delta x=1$.
 
The wall thickness, defined by
\begin{equation}
{\omega}_0={{\pi \phi_0} \over {\sqrt {2 V_0}}}
\label{thick}
\end{equation}
is set to be equal to $5$. The kinetic energy of the field $\phi$ is 
calculated by
\begin{equation}
E_{kin}=\frac{1}{8\pi} \sum_{i,j} {\dot \phi}_{i,j}\, .
\label{kinetic}
\end{equation}
On the other hand, the rest energy of the walls is calculated
by multiplying the 
comoving area of the walls, A, by the energy density per comoving area, 
which can be written as
\begin{equation}
\sigma=\frac{2 \gamma_w V_0 \omega_0}{3 \pi^2}.
\label{sigma}
\end{equation}
with $\gamma_w=(1-v_w^2)^{-1/2}$, and $v_w$ is the value of the 
physical velocity of the 
domain walls. Finally the total area of the walls is defined as the area 
of the surfaces on which $\phi=0$ and is computed using the method 
described in ref. \cite{PRS}.

%%%%%%%%%%%%%%%%%%%%%%%%%%%%%%%%%%%%%%%%%%%%%%%%%%%%%%%%%%%%%%%%%%%%%%%%%%
\section{Results and discussion}
\label{redi}

As pointed out above, it will be of fundamental importance to study
the dynamics of the wall network at late times, as the Hubble length
becomes larger than the typical size 
of the major axis of a domain wall. A crucial issue will be the timescale
required for the wall network to switch from the non-relativistic
regime to the relativistic one. For our present purposes, the main difference
between these two regimes is that a friction-dominated network can
remain anisotropic if it froze out that way, whereas a relativistic network
will rapidly become isotropic and erase any imprints from the
earlier anisotropic phase. In our simulations we ignore the possibility that 
the network can be friction dominated due to particle scattering \cite{ms3}
when the domain walls come back inside 
the horizon---again, this would only enhance the effects we are describing.

\begin{figure*}[t]
\vskip0in
%\vbox{\centerline{
%\epsfxsize=0.5\hsize\epsfbox{box1.ps}}
%\vskip0in}
%\vbox{\centerline{
%\epsfxsize=0.5\hsize\epsfbox{box2.ps}}
%\vskip0in}
\caption{The same physical size of the same domain wall simulation. The 
bottom one however has been stretched along the y direction by a factor of 2. 
The horizon size of the top box is 1/8 of the linear size of the box shown,
and this is itself only a  fraction 1/16 of the side of the
whole simulation box.]}
\label{fig_boxes}
\end{figure*}

We consider three simulations with different initial conditions. 
In the first one (case I) we evolve the initial network generated in the 
manner specified in the previous section from the conformal time $\eta_i=1$.
In the second one (case II) the initial conditions at the time $\eta_i=1$ 
were specified by the network configuration of the previous simulation 
at the conformal time $\eta_*=20$, with the velocities reset to zero.
Physically, this corresponds to starting with the network
outside the horizon. Finally, the case III is 
similar to the second one but with the initial network of case II 
stretched in the $y$ direction by a factor of $E=2$ (see fig. \ref{fig_boxes}),
and corresponds to the anisotropic case.
We have performed $1024^2$ simulations for each of the three cases, plus
an additional $2048^2$ run of case I, in order to test for possible box effects.

\begin{figure*}[t]
\begin{center}
\leavevmode \epsfbox{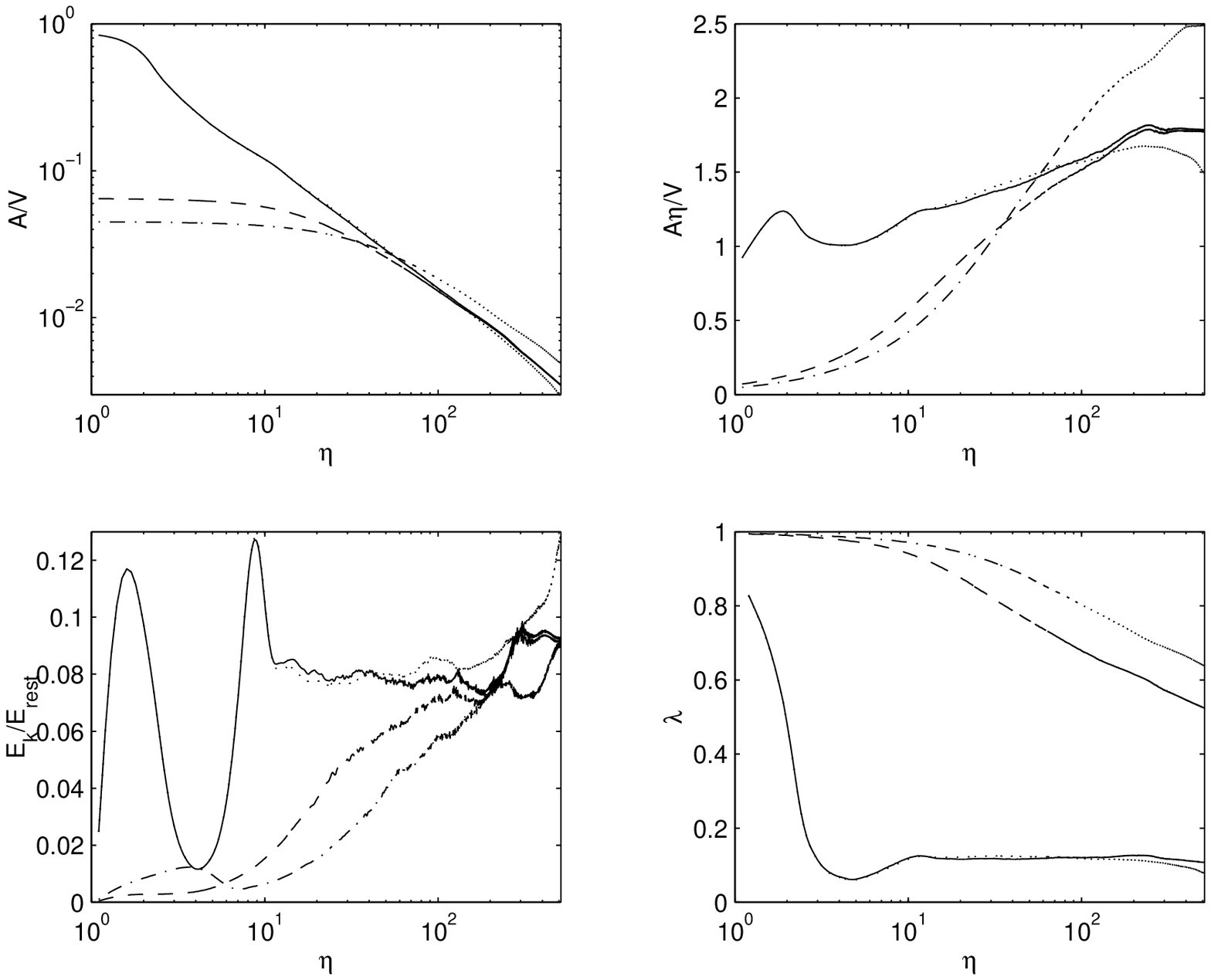}\\
\end{center}
\vskip1in
\caption[]{Evolution of several properties  of a domain
wall network as a function of the conformal time $\eta$ during the matter era. 
Here, $E_{k}$ and $E_{rest}$ are the kinetic and rest energies associated 
with the scalar filed $\phi$, $A /V$ is the comoving wall area per unit 
comoving volume of the two-dimensional simulations, $\lambda$ is a 
`scaling coefficient' defined by $\lambda=d \ln (A \eta/V) / d \ln \eta$ and 
$\eta$ is the conformal time. The solid, dashed and dash-dotted curves 
correspond to $1024^2$ simulations with different initial conditions. 
Cases I and II have isotropic initial conditions but in case two the size of 
the domain walls becomes comparable with horizon only at a conformal time 
$\eta_*=20$. Case III is similar to case II but with the initial network 
of case II stretched in the $y$ direction by a factor of $E=2$ (see text).
The dotted curve corresponds to a $2048^2$ run of case I.}
\vskip0.5in
\label{figure3}
\end{figure*} 

For each run we plot (see fig. \ref{figure3}) the ratios $A/V$ and $A\eta/V$
(note that $A$ and $V$ are
the {\em comoving} area and volume, respectively), as well as the ratio
of the kinetic and rest energies, as in Press, Ryden and Spergel \cite{PRS}.
These are plotted from the beginning of the simulation until the time
when the horizon becomes one half (for the $1024^2$ runs) or one quarter
(for the $2048^2$ run) of the box size.
In addition to these (which we plot mainly for the purposes of comparison with
previous work \cite{PRS}) we plot a `scaling coefficient' which will
be our main analysis tool. We will define it by analogy with
the cosmic string case \cite{ms1,ms3}, as follows.
Assume that
\begin{equation}
\eta\frac{A}{V}\propto\eta^\lambda\, ;
\label{coefficient}
\end{equation}
then what we plot is the `instantaneous' or `effective' value of $\lambda$ as
a function of conformal time. For a given $\lambda$, the {\em physical}
network correlation $L$ length will be evolving as
\begin{equation}
L\propto t^{1-\lambda/3}\, , \qquad \frac{L}{a}\propto \eta^{1-\lambda}\, .
\label{evolllambda}
\end{equation}

Note that $\lambda$ can, in general, be a
time-dependent quantity. However, for the two scaling regimes discussed above,
we expect it to be a constant, namely
\begin{equation}
\lambda_{nr}=1\,
\label{lamstretch}
\end{equation}
in the non-relativistic limit where the network is being conformally
stretched, and
\begin{equation}
\lambda_{r}=0\,
\label{lamscal}
\end{equation}
in the linear scaling regime.

From these it is trivial to deduce the behaviour of $A/V$ and $A\eta/V$
in both scaling regime. One expects
\begin{equation}
\eta\frac{A}{V}\propto\eta\, \qquad \frac{A}{V}\propto const.
\label{coefficient11}
\end{equation}
in the non-relativistc regime and
\begin{equation}
\eta\frac{A}{V}\propto const\, \qquad \frac{A}{V}\propto\eta^{-1}
\label{coefficient22}
\end{equation}
in the linear scaling regime.
Similarly, the ratio $E_{k}/E_{rest}$ should be a constant in the linear
scaling regime (with its numerical value providing a measure of the
characteristic network scaling speed), and it should approach zero in
the non-relativistic limit.

Firstly, we note that the two case I runs produce very similar results:
significant differences can only be seen at late times. This is an indication
that the resolution we are using is adequate for our present purposes.
As expected, the network in case I becomes relativistic very quickly, while
those of cases II and III start in the extreme non-relativistic regime and
only evolve away from it fairly slowly, after they re-enter the horizon.

More importantly, there are two non-trivial observations to be made.
Firstly, we confirm that there is a correction to the linear scaling regime.
We find
\begin{equation}
\lambda_{sc}\sim0.12\, ,
\label{lamscalnew}
\end{equation}
which corresponds to
to evolve in a linear scaling regime where
\begin{equation}
L\propto t^{0.96}\, , \qquad \rho_w\propto t^{-0.96}\, ,
\label{newscaling}
\end{equation}
in agreement with the previous result by Press, Ryden and Spergel \cite{PRS}.
This means that the network is not straightening out as fast as allowed
by causality.
Secondly, the rates at which the networks in cases II and III approach
the relativistic regime are different. One might expect this on physical
grounds: if the network is stretched in one direction, then there are in
fact different `network correlation lengths' for each direction, and
interactions between the domain walls will tend to occur faster along
the directions with smaller correlation lengths, and more slowly
in the others.

Another way of saying this is that the network will only start evolving
towards the relativistic regime when its larger axis has re-entered the
horizon. Note that this mechanism also tends to make the domain wall
network more isotropic. So one can naively say that the approach to the linear
scaling regime takes longer in an anisotropic universe because the dynamics
of the walls must accomplish two tasks (make the wall network relativistic
and isotropic) rather than just one.

%%%%%%%%%%%%%%%%%%%%%%%%%%%%%%%%%%%%%%%%%%%%%%%%%%%%%%%%%%%%%%%%%%%%%%%%%%

\section{Conclusion}
\label{conc}

In this paper we have discussed a simple example of what we believe to be
a rather generic feature of topological defect models, namely that they
can easily retain information about the properties of the very early universe.
This information is encoded in the scaling (ie, `macroscopic') and
statistical (ie, `microscopic') properties of the defect networks.
This is even more relevant given the fact that defects can survive significant
amounts of inflation. Hence, they can provide a unique probe of
the pre-inflationary universe. The two crucial scales in the problem are the
defect mass scale and the epoch when the defects come back inside the
horizon.

Specifically, we have discussed the role of domain walls. We have highlighted
the existence of two scaling regimes for the domain wall network, in agreement
with previous work \cite{PRS}. Furthermore, we have shown that an anisotropic
network re-entering the horizon will take longer to approach scaling than an
isotropic one. Hence, if the very early universe had an anisotropic phase
which was erased by an inflationary epoch, and if domain walls are present,
then the walls can retain an imprint of the earlier phase, and this can have
important observational consequences, eg for structure formation scenarios.

As is well known, there are quite strong constraints \cite{zeld,vsh} on the
mass of domain walls formed after inflation. These are basically due to the
fact that their density will decay more slowly than the radiation and
matter densities. However, essentially all of these can be evaded (or at least
significantly relaxed) by walls forming before or during inflation (and also
by walls evolving in a friction-dominated regime).
Having said this, how could these anisotropies be detected?
The most naive answer would be through their imprint on CMB, but this is only
true if their energy density is not too low, and such models are constrained
in a variety of other ways (not only from the cosmology side, but also from
the high-energy physics side). The case of `light' walls is therefore more
interesting: note that just like in the case of `light strings' \cite{ms3},
these are expected to be friction-dominated throughout most of the cosmic
history. Here the observational detection of the effects
we have described becomes somewhat non-trivial.
The best way of doing it should be through observations of numbers of
objects as a function of redshift in different directions (assuming that one
has a reliable understanding of other possible evolutionary effects).
Two specific examples would be large-scale velocity flows \cite{zeh}
and gravitational lensing statistics of extragalactic surveys \cite{quast}.

Finally, there is also an important implication of our work
if at least one of the minima of the
scalar field potential has a non-zero energy density, which is an anisotropic
non-zero vacuum density. In a subsequent, more detailed publication, we shall
discuss this scenario in more detail, as well as the analogous one for
cosmic strings.

To conclude, we have shown that the importance of topological defects as a
probe of cosmological physics goes well beyond structure formation.
Even if defects turn out to be unimportant for structure formation they can 
still (if detected) provide us with extremely valuable information about
the physical conditions of the very early universe.

%%%%%%%%%%%%%%%%%%%%%%%%%%%%%%%%%%%%%%%%%%%%%%%%%%%%%%%%%%%%%%%%%%%%%%%%%%
\acknowledgements

We would like to thank Paulo Carvalho for enlightening discussions.
C.M. is funded by JNICT (Portugal) under
`Programa PRAXIS XXI' (grant no. PRAXIS XXI/BPD/11769/97).

%%%%%%%%%%%%%%%%%%%%%%%%%%%%%%%%%%%%%%%%%%%%%%%%%%%%%%%%%%%%%%%%%%%%%%%%%%


\begin{references}
\bibitem{kolb}
E.W. Kolb \& M.S. Turner, {\it The Early Universe}, (Addison-Wesley, 1994).
\bibitem{guth}
A.H. Guth, {\it Phys.\ Rev.\ } {\bf D23}, 347 (1981).
\bibitem{linde}
A.D. Linde, {\it Particle Physics and Inflationary Cosmology} (Harwood, Chur,
Switzerland, 1990).
\bibitem{lyth}
A.R. Liddle \& D.H. Lyth, {\it Phys.\ Rep.\ } {\bf 231}, 1 (1993);

D.H. Lyth \& A. Riotto {\it Phys.\ Rep.\ } {\bf 314}, 1 (1999).
\bibitem{curv}
S. Gratton, T. Hertog \& N.G. Turok, {\it astro-ph/9907212} (1999).
\bibitem{vsh}
A. Vilenkin \& E. P. S. Shellard, {\it Cosmic Strings and other Topological
Defects}, (Cambridge University Press: Cambridge, 1994).
\bibitem{shafi}
Q. Shafi \& A. Vilenkin, {\it Phys.\ Rev.\ Lett.\ } {\bf 52}, 691 (1984);

Q. Shafi \& A. Vilenkin, {\it Phys.\ Rev.\ } {\bf D29}, 1870 (1984).
\bibitem{yoko}
J. Yokoyama, {\it Phys.\ Lett.\ } {\bf B212}, 273 (1988);

J. Yokoyama, {\it Phys.\ Rev.\ Let.\ } {\bf 63}, 712 (1989).
\bibitem{hodges}
H.M. Hodges \& J.R. Primack, {\it Phys.\ Rev.\ } {\bf D43}, 3155 (1991).
\bibitem{acm2}
P.P. Avelino, R.R. Caldwell \& C.J.A.P. Martins,
{\it Phys.\ Rev.\ } {\bf D59}, 123509 (1999).
\bibitem{ms1}
C.J.A.P. Martins \& E.P.S. Shellard, Phys. Rev. {\bf D53}, 575 (1996);

C.J.A.P. Martins \& E.P.S. Shellard, Phys. Rev. {\bf D54}, 2535 (1996).
\bibitem{ms2}
P.P. Avelino, R.R. Caldwell \& C.J.A.P. Martins, Phys. Rev. {\bf D56}, 4568 (1997).
\bibitem{ms3}
C.J.A.P. Martins, {\em Quantitative String Evolution}, Ph.D. Thesis,
University of Cambridge (1997).
\bibitem{bgt}
M. Bucher, A.S. Goldhaber \& N.G. Turok, {\it Phys.\ Rev.\ } {\bf D52}, 3314 (1995).
\bibitem{vil}
A. Vilenkin, {\it Phys.\ Rev.\ } {\bf D56}, 3258 (1997).
\bibitem{zeld}
Ya.B. Zel'dovich, I. Kobzarev \& L.B. Okun, {\it Soviet Phys.\ JETP},
{\bf 40}, 1 (1975).
\bibitem{PRS} 
W.H. Press, B.S. Ryden \& D.N. Spergel, {\it Ap.\ J.\ } {\bf 347}, 590 (1994).
\bibitem{anis}
M. Carmeli, Ch. Charach \& S. Malin, {\it Phys.\ Rep.\ } {\bf 76}, 79 (1981);

J. Wainwright \& G.F.R. Ellis, {\it Dynamical Systems in Cosmology} (Cambridge
University Press, 1997).
\bibitem{eps}
E.P. Shellard, {\it private communication} (1999).
\bibitem{zeh}
I. Zehavi \&  A. Dekel, astro-ph/9904221 (1999).
\bibitem{quast}
R. Quast \& P. Helbig, {\it  Astron.\ Astroph.\ } {\bf 344}, 721 (1999).

\end{references}
\end{document}